\documentclass[aps,prl,twocolumn,showpacs,superscriptaddress,email]{revtex4-1}  
\usepackage[pdftex]{graphicx}  
\usepackage{dcolumn}   
\usepackage{bm}        
\usepackage{amssymb}   

\usepackage{amsmath}
\usepackage[latin1]{inputenc}

\usepackage{color}

\newcommand{\ket}[1]{\left| #1\right\rangle}

\begin{document}

\title{Entanglement of two qubits mediated by one-dimensional plasmonic waveguides}

\author{A. Gonzalez-Tudela}
\affiliation{Departamento de F\'isica Te\'orica de la Materia
Condensada, Universidad Aut\'onoma de Madrid, Madrid 28049, Spain}

\author{D. Martin-Cano}
\affiliation{Departamento de F\'isica Te\'orica de la Materia
Condensada, Universidad Aut\'onoma de Madrid, Madrid 28049, Spain}

\author{E. Moreno}
\affiliation{Departamento de F\'isica Te\'orica de la Materia
Condensada, Universidad Aut\'onoma de Madrid, Madrid 28049, Spain}

\author{L. Martin-Moreno}
\affiliation{Instituto de Ciencia de Materiales de Aragon (ICMA)
and Departamento de Fisica de la Materia Condensada,
CSIC-Universidad de Zaragoza, E-50009 Zaragoza, Spain}

\author{C. Tejedor}
\email[Corresponding author: ]{carlos.tejedor@uam.es}
\affiliation{Departamento de F\'isica Te\'orica de la Materia
Condensada, Universidad Aut\'onoma de Madrid, Madrid 28049, Spain}

\author{F.J. Garcia-Vidal}
\email[Corresponding author: ]{fj.garcia@uam.es}
\affiliation{Departamento de F\'isica Te\'orica de la Materia
Condensada, Universidad Aut\'onoma de Madrid, Madrid 28049, Spain}

\begin{abstract}
We investigate qubit-qubit entanglement mediated by plasmons supported by one-dimensional waveguides. We explore both the situation of spontaneous formation of entanglement from an unentangled state and the emergence of driven steady-state entanglement under continuous pumping. In both cases, we show that large values for the concurrence are attainable for qubit-qubit distances larger than the operating wavelength by using plasmonic waveguides that are currently available. 
\end{abstract}
\pacs{42.50.Ex, 03.67.Bg, 73.20.Mf, 42.79.Gn}
\date{\today}
\maketitle














As a direct consequence of the quantum superposition principle, a system composed of subsystems has states that cannot be factorized in products of states of its components. This non-separability, labeled as entanglement, is at the heart of quantum cryptography, quantum teleportation, or other two-qubit quantum operations \cite{nielsen,haroche}. Exploited at first in systems like optics, atoms, or ions, entanglement is becoming more and more attainable in condensed matter physics. In particular, short distance entanglement is now available for spin or charge degrees of freedom in quantum dots (QDs), nanotubes, or molecules \cite{makhlin,kouwenhoven07,weber10,defranceschi}. However, for transmission of information at long distances, large separations between the components are needed. For this purpose, correlation among the two qubits must be mediated by virtual bosons. Photons, either in the range of microwaves for coupling superconducting qubits \cite{majer07a} or in the visible range for QDs \cite{imamoglu99a,gallardo10a,laucht10a}, molecules, or NV centers in diamond \cite{dutt07}, are the usual candidates to play this role. 

Here we investigate a feasible proposal for long-distance entanglement of two qubits by using plasmons instead of photons.  We consider the plasmon-polariton modes supported by one-dimensional (1D) plasmonic waveguides (PWs), see top panel of Fig.1. PWs have been studied during the last years as promising candidates to build up a new kind of photonic circuitry \cite{Ebbesen08}. The propagating plasmons associated with these structures are characterized by both a subwavelength light confinement and long enough propagation lengths \cite{Moreno08}. Coupling between quantum emitters and PWs has been also addressed \cite{Chang06a,Akimov07a}. These works show that the $\beta$-factor, which measures the fraction of the emitted radiation that is captured by the propagating mode, can be close to $1$ in realistic PWs. This is due to the subwavelength nature of the plasmon field in a 1D-PW. Very recently, these large $\beta$-factors have been exploited to modulate the energy transfer and superradiance phenomena appearing when two quantum emitters are placed at 1D-PWs like channel or wedge structures \cite{Martin-Cano10}. In this Letter we show that PWs can also be used to obtain a large degree of entanglement in two qubits separated by distances larger than the operating wavelength.

\begin{figure}[htbp]
\begin{center}
\includegraphics[width=0.99\linewidth,angle=0]{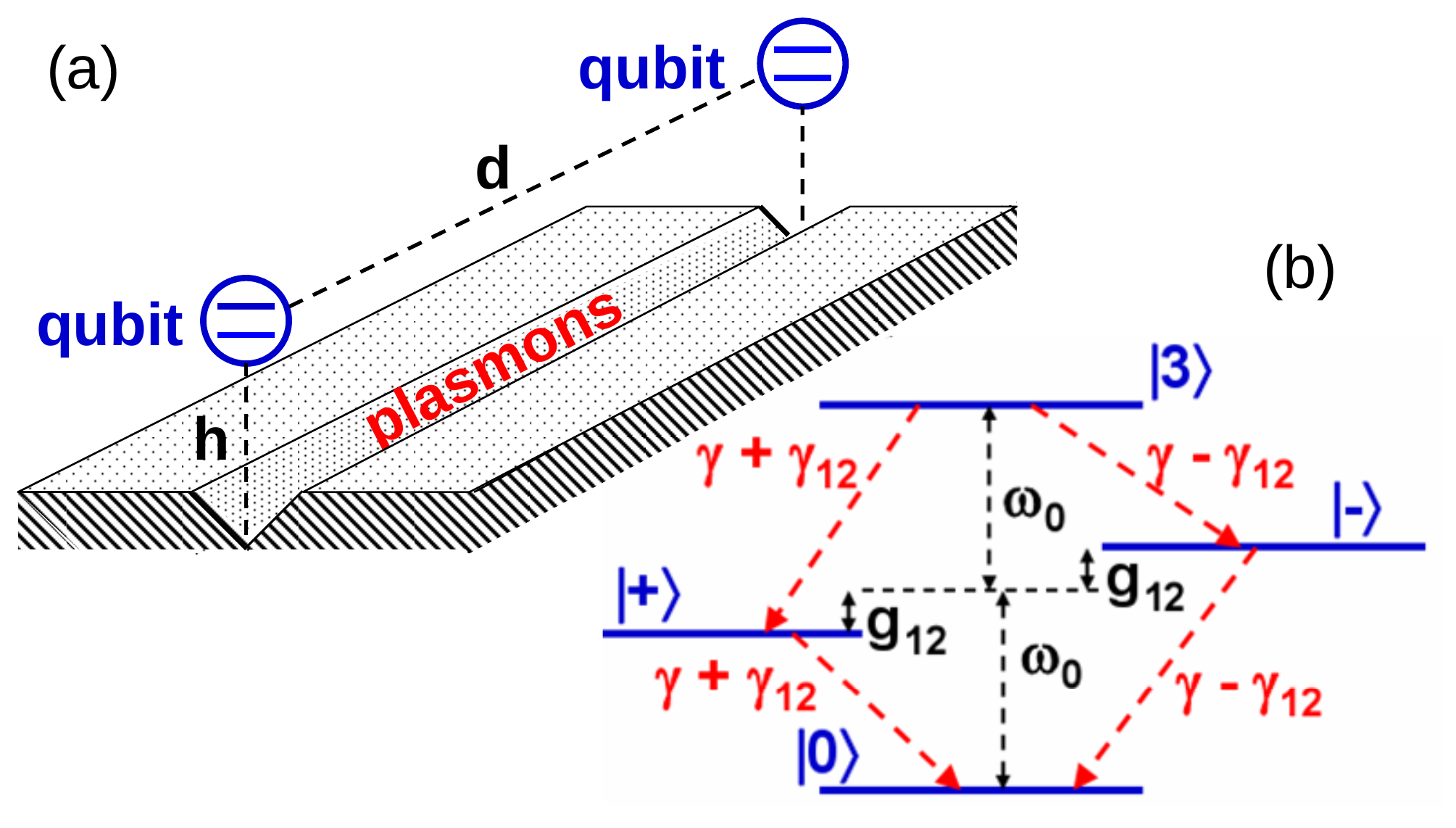}
\end{center}
\caption{(Color online) (a) Two qubits interacting with a plasmonic waveguide, in this case a channel waveguide. (b) Scheme of levels, couplings, and decays in the particular case where $\omega_{1}=\omega_{2}=\omega_0$ and $\gamma_{11}=\gamma_{22}=\gamma$.}
\label{scheme}
\end{figure}
%

The dynamics of the density matrix $\rho$ for two qubits is described, after tracing out over the degrees of freedom of the plasmons, by a master equation \cite{ficek02a,dzsotjan10a}

\begin{equation}
\label{masterequation}
\partial_t \rho=\frac{i}{\hbar}[\rho,H]+\sum_{i,j=1,2}\frac{\gamma_{ij}}{2}(2 \sigma_i \rho \sigma_j^\dagger- \sigma_i^\dagger\sigma_j\rho-\rho\sigma_i^\dagger\sigma_j),
\end{equation}
where $\sigma_i^\dagger,\sigma_i$ are the raising and lowering operators for each qubit. The ingredients of Eq. (1) are determined by the classical Green's function describing the electromagnetic interaction between two dipole moments, $\mu_{1}$ and $\mu_{2}$, placed at locations ${\bf r}_1$ and 
${\bf r}_2$, ${\bf G}(\omega,{\bf r}_1,{\bf r}_2)$. For two qubits with the same characteristic frequency $\omega_0$, the hamiltonian can be written as

\begin{equation}
\label{hamilt}
H=\hbar \omega_0 \sum_{i=1,2} \sigma_i^\dagger\sigma_i+ g_{12}(\sigma_1^\dagger\sigma_2+\sigma_2^\dagger\sigma_1).
\end{equation}

The coherent part of Eq.(1) reflects the effective interaction between the qubits that is provided by the exchange of virtual bosons\cite{dzsotjan10a}:

\begin{equation}
\label{lambshift}
g_{12}=\frac{1}{\pi\epsilon_0 \hbar}\mathcal{P}\int_0^\infty \frac{\omega^2 Im[\mu_1^{*} {\bf G}(\omega,{\bf r}_1,{\bf r}_2)\mu_2]}{c^2(\omega-\omega_0)}d\omega,
\end{equation}
whereas the rates of the non-coherent terms are given by

\begin{equation}
\label{decay}
\gamma_{ij}=\frac{2 \omega_0^2}{\epsilon_0 c^2 \hbar}Im[\mu_i^{*} {\bf G}(\omega_0,{\bf r}_i,{\bf r}_j)\mu_j],
\end{equation}
with $i,j=1,2$ and $\gamma_{12}=\gamma_{21}$.  Equations (3) and (4) involve a point-dipole emitter approach, which is accurate enough for qubits such as atoms, small molecules or NV centers in diamond. For big molecules or QDs with sizes of a few tens of nanometers, a more realistic description of the quantum emitter is usually required \cite{andersen10}. 

When the plasmon supported by the PW is the dominant decay channel (i.e., large $\beta$-factor), a very good approximation for the total Green's function can be obtained by only considering its plasmon contribution, ${\bf G} (\omega,{\bf r}_1,{\bf r}_2) \approx {\bf G}_{pl} (\omega,{\bf r}_1,{\bf r}_2)$ \cite{Martin-Cano10}. In this way, analytical expressions for both $g_{12}$ and $\gamma_{12}$ can be easily derived: 
 
\begin{eqnarray}
\label{coupling-crossdecay}
g_{12} & = & \frac{\gamma}{2} \beta e^{-d/(2 L)} \sin(k_{pl} d ) \nonumber \\
\gamma_{12} & = & \gamma \beta e^{-d/(2 L)} \cos(k_{pl} d),
\end{eqnarray}
where  $k_{pl}$ and $L$ are the wavenumber and propagation length of the plasmon, respectively. These two magnitudes, $k_{pl}$ and $L$, depend on the operating frequency,$\omega_0$. In deriving Eq.(5) we have assumed that the two qubits are equal and are placed at two equivalent positions along the waveguide, such that $\gamma_{11}=\gamma_{22}=\gamma$, and separated by a distance $d$. We define the modal wavelength of the plasmon, $\lambda_{pl}$, as $\lambda_{pl}=2\pi/k_{pl}$. The crucial point of Eq.(\ref{coupling-crossdecay}) is the $\pi/2$ phase shift between the coherent and incoherent parts of the coupling, which allows switching off one of the two contributions while maximizing the other by just choosing the inter-qubit distance. This opens the possibility of modulating the degree of entanglement.  

To test the feasibility of our proposal, we have carried out extensive numerical calculations on a particular PW, a V-groove milled on a silver film. In Fig. 2(a) we render the dispersion relation (energy versus wavenumber) of the propagating plasmon supported by a V-groove. This type of plasmon are usually called channel plasmon polaritons (CPPs). The geometrical parameters are taken from realistic structures: the angle of the groove is $20^0$ and its height is $140$ nm,  but similar results would be obtained for other sets of parameters. The evolution of the propagation length, $L$, of the CPP with the operating wavelength, $\lambda$, is shown in the inset of Fig. 2(a). As expected, $L$ increases as $\lambda$ is enlarged. The dependence of the $\beta$-factor with both $\lambda$ and the vertical distance of the qubit(s) to the apex of the V-groove (see Fig.1(a)) is displayed in Fig. 2(b). Importantly, $\beta$- factors larger than $0.9$ are attainable for a broad range of $\lambda$'s and within a large spatial region. Let us remark that the best $\beta$-factors attained with dielectric waveguides are typically much lower than those obtained with PWs (for instance, $\beta$-values not higher than $0.5$ are reported in \cite{Martin-Cano10} for a GaAs fibre of $50$ nm radius).

\begin{figure}
\begin{center}
\includegraphics[width=0.99\linewidth,angle=0]{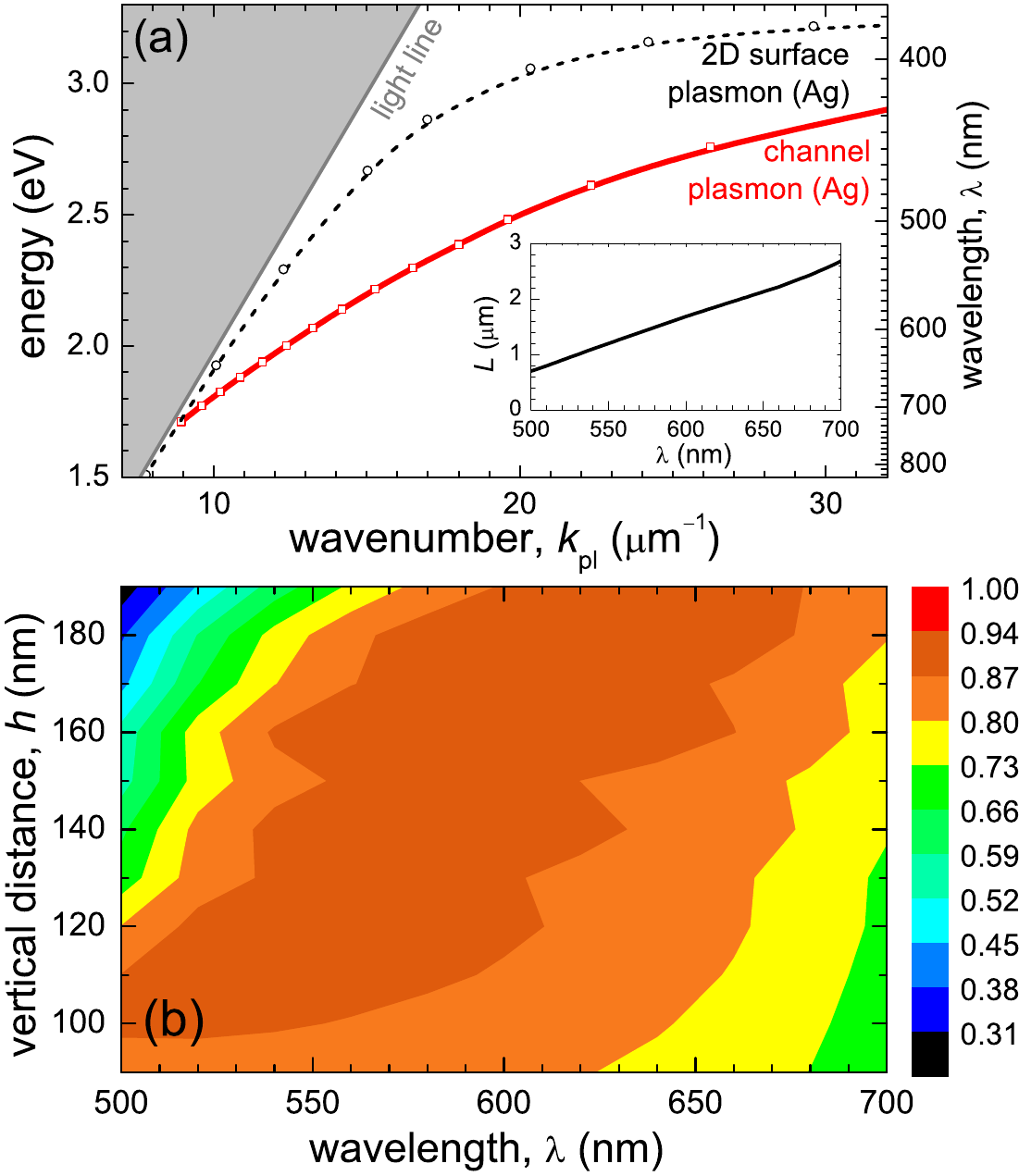}
\end{center}
\caption{(Color online) (a) Dispersion relation (red curve) of a CPP mode supported by a V-groove of angle $20^0$ and height $140$ nm. For comparison we show the dispersion of a surface plasmon mode of an infinite 2D silver surface (black dotted line). The inset displays the wavelength dependence of the propagation length, $L$. (b) $\beta$-factor versus wavelength, $\lambda$, and vertical distance, $h$, associated with the CPP mode.}
\label{beta-L}
\end{figure}

Once we have introduced the ingredients of Eq. (1), now we discuss how this equation is solved. 
The most adequate basis to represent the dynamics of Eq.(\ref{masterequation}) is the one depicted in Fig. 1(b): \{$\ket{0}=\ket{g_1,g_2},\ket{\pm}=\frac{1}{\sqrt{2}}(\ket{e_1,g_2}\pm\ket{g_1,e_2}),\ket{3}=\ket{e_1,e_2}$\}, where $g_i/e_i$ labels the ground/excited state of the $i$-qubit. Depending on both the sign and absolute value of  $\gamma_{12}$, one of the states $\ket{\pm}$ can be practically decoupled from the dynamics of the rest of states. Once the density matrix $\rho (t)$ is obtained by numerically solving Eq. (\ref{masterequation}), the entanglement of the two qubits is quantified by means of the concurrence, $C$, defined as proposed by Wootters \cite{wootters98a}. The two main ingredients controlling the dynamics of the two qubits  ($g_{12}$ and $\gamma_{12}$) affect $C$ in very different ways.  The coherent coupling $g_{12}$ produces oscillations, whereas the cross-decay term, $\gamma_{12}$, produces a non-oscillatory contribution to $C$. These two effects are discussed below in two different situations.  First, we analyze the case in which the system is initially prepared in a given unentangled state from which it decays spontaneously. In the second situation, the two qubits are continuously pumped by an external laser to reach a stationary state.

\begin{figure}
\begin{center}
\includegraphics[width=0.99\linewidth,angle=0]{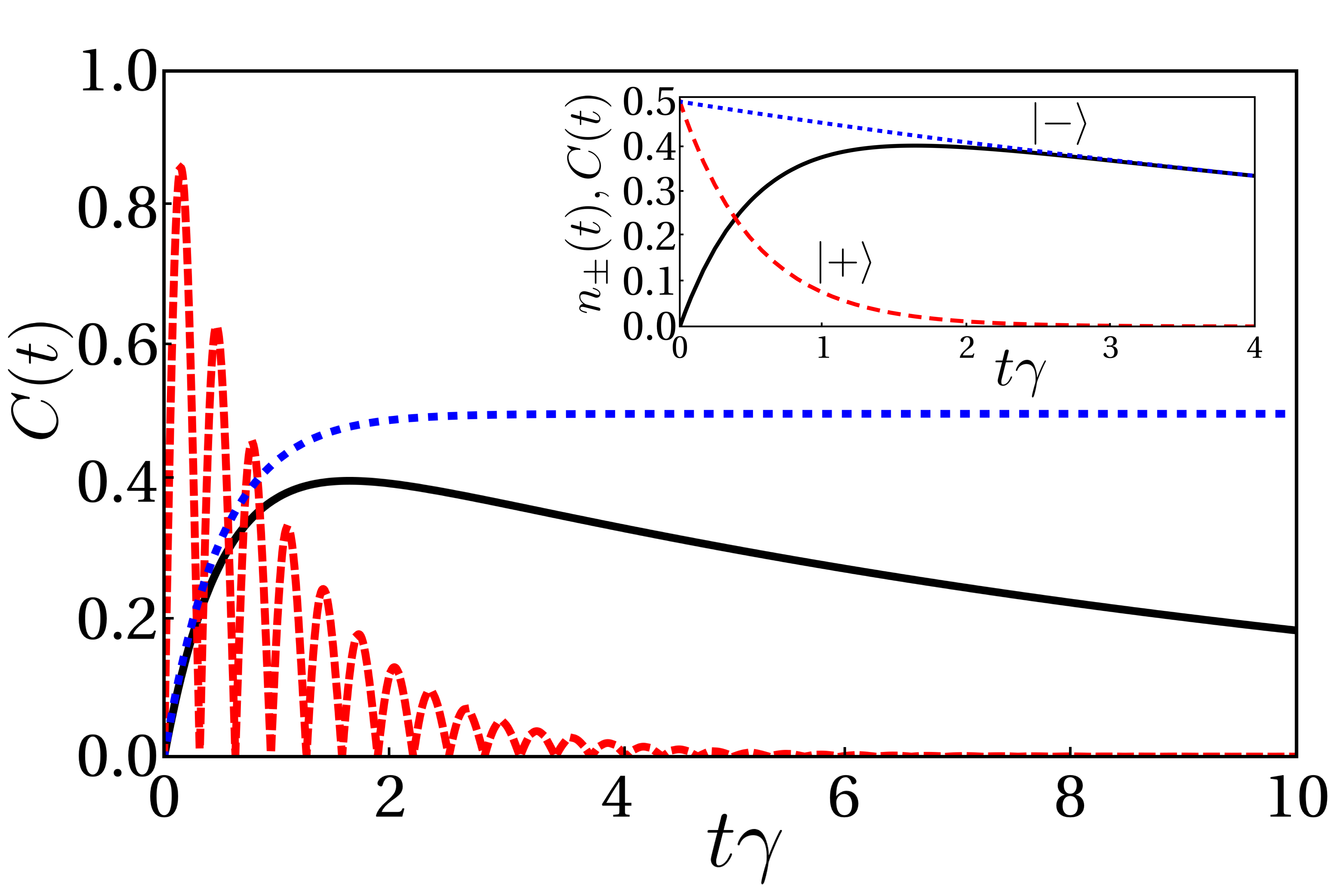}
\end{center}
\caption{(Color online) Concurrence as a function of time when just one of the two qubits is initially excited. Continuous black line corresponds to two qubits entangled by means of a channel PW with $\overline{\beta}=0.9$, $k_{pl} d=2 \pi$. Dashed red line is for two qubits in a cavity with a detuning such that it produces $g_{12}/\gamma=5$ and $\gamma_{12}=0$. Dotted blue line is for the ideal case of both PW and CQED (see text). The inset shows the time evolution of $C$ and the populations of states $\ket{+}$ and $\ket{-}$ for the non-ideal PW case.}
\label{spontaneous}
\end{figure}

In order to analyze the spontaneous formation of entanglement, one can initially prepare the system in the $\ket{e_1,g_2}$ state. In this case, the concurrence takes the form

\begin{eqnarray}
C(t)=\sqrt{[\rho_{++}(t)-\rho_{--}(t)]^2+4Im [\rho_{+-}(t)]^2},
\label{c(t)}
\end{eqnarray}
where $\rho_{\pm \pm}$ are density matrix elements in the basis $\ket{\pm}$. The dynamics of $C$ is shown in Fig. \ref{spontaneous} for three different situations. With a solid black  curve we render $C(t)$ for the case of a channel PW where $\gamma_{12}$ and $g_{12}$ are given by Eq. (\ref{coupling-crossdecay}), with $\bar{\beta} \equiv \beta e^{-d/(2L)}=0.9$ and $k_{pl} d=2\pi$. The concurrence is characterized by a fast initial increase followed by a very slow decay. For this case, the coherent oscillations produced by $g_{12}$ are completely quenched ($g_{12}=0$) and the cross-decay term ($\gamma_{12}$) dominates. This dynamics can be easily understood from the time evolution of the populations of the two entangled states $\ket{\pm}$ (see inset of Fig. 3). These states are equally populated initially but the decay of the state $\ket{+}$ is very fast ($\gamma+\gamma_{12}$) while the decay of $\ket{-}$ is very slow ($\gamma-\gamma_{12}$). The asymmetry between the two cascades is responsible for the long lifetime of $C$ while the imbalance among the populations of states $\ket{\pm}$ determines $C$ due to the first term in Eq.(\ref{c(t)}). With a dotted blue line we represent the {\it ideal} case that corresponds to $\bar{\beta}=1$  ($\beta=1$ and $L=\infty$) and $k_{pl} d=2\pi$. In this case, the concurrence tends asymptotically to a steady state value of $0.5$. 

It is worth comparing our PW-based entanglement with other schemes for achieving large entanglement that have been proposed before. In particular, embedding two qubits in a photonic cavity (CQED) \cite{imamoglu99a} offers many possibilities for controlling the photon emission. As entanglement in CQED only relies in the coherent term, $\omega_0$ must be tuned to the frequency of the cavity mode. In the case of perfect tuning \cite{ficek02a}, the evolution of $C$ with time is determined by $\gamma_{12}=g^2/\gamma$, $g$ being the qubit-cavity coupling. The time-dependent $C$ becomes equal to that of an ideal PW (dotted blue curve in Fig. 3). However, in realistic implementations of CQED, one must work with a detuning $\Delta$ that is comparable to $g$. By using a Schrieffer-Wolf transformation one can obtain a master equation like Eq.(\ref{masterequation}) with $g_{12}=g^2/\Delta$ and $\gamma_{12}\approx 0$. In Fig. 3 we render a simulation to illustrate CQED-based entanglement (see red dashed curve) in which we have used $g_{12}/\gamma=5$ and $\gamma_{12}=0$ as realistic values taken from experiments \cite{gallardo10a,laucht10a}. Although in this case $C$ can be significantly larger than the one obtained with realistic PWs, this only occurs in a very short time scale and presents very fast oscillations. Another advantage of the use of PWs is that they enable the emergence of long-distance qubit-qubit correlations, as we show below. 

\begin{figure}
\begin{center}
\includegraphics[width=0.99\linewidth,angle=0]{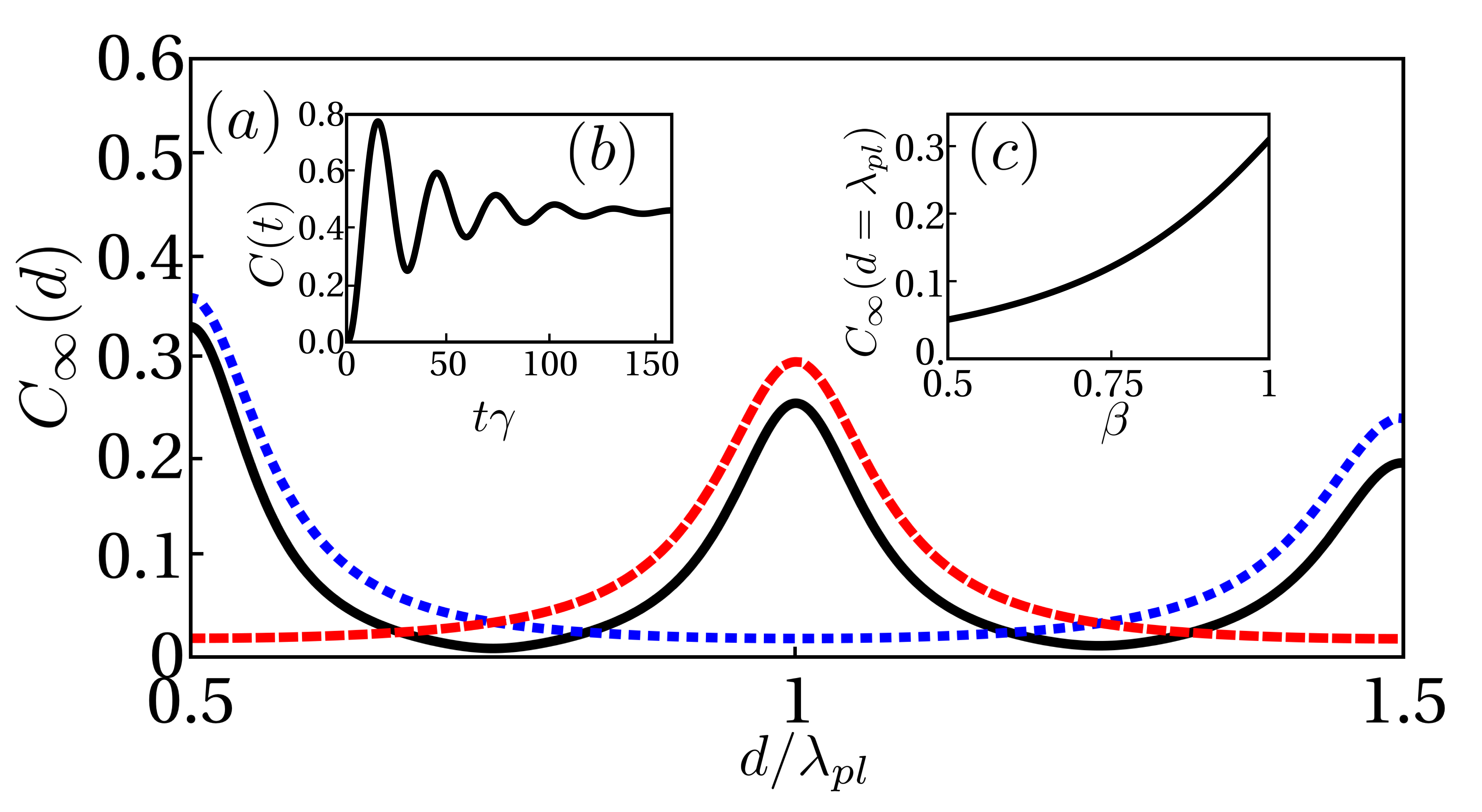}
\end{center}
\caption{(Color online) Panel (a) shows the steady state concurrence as a function of the separation between two equal qubits for $\beta=0.94$, $L=2 \, \mu$m and three different laser configurations: $\Omega_1=0.15\gamma$ and $\Omega_2=0$ (solid black curve), $\Omega_1=\Omega_2=0.1\gamma$ (dotted blue line) and $\Omega_1=-\Omega_2=0.1\gamma$ (dashed red line). Insets (b) and (c) correspond to the first type of pumping $\Omega_1=0.15\gamma$ and $ \Omega_2=0$. Panel (b) shows the concurrence time-evolution for a system with $\overline{\beta}=0.99$, while the $\beta$-dependence of the steady-state concurrence evaluated at $d=\lambda_{pl}$ is rendered in panel (c).}
\label{stationary-laser}
\end{figure}

A continuous pumping is required in order to have a stationary state with a high degree of entanglement. For sufficiently separated qubits, the stationary state can be modulated by acting independently and resonantly($\omega_{\mathrm{laser}}=\omega_0$ on each qubit with a laser beam of Rabi frequency $\Omega_i$. A new term, $\sum_i \hbar \Omega_i(\sigma^\dagger_i+\sigma_i)$ must be included in the hamiltonian given by Eq.(\ref{hamilt}). When the system is initially prepared at state $\ket{0}$, one can apply the continuous laser only on qubit $1$. Inset (b) of Fig. \ref{stationary-laser} shows, for an almost ideal PW and very short distances, the transient dynamics of the concurrence when $d=\lambda_{pl} /2 $ so that $\gamma_{12}$ is present while $g_{12}$ is quenched. At the beginning, clear oscillations are observed with the concurrence becoming larger than $0.5$ for some time intervals, $\Delta t \simeq 10/\gamma$. 

More interesting is the discussion of the stationary concurrence, $C_\infty$, for different laser intensities and separations between qubits. Here we consider realistic values for $\beta$ and $L$ taken from Fig. 2(b): $\beta=0.94$ and $L=2 \, \mu$m, which correspond to a wavelength of around $640$ nm and vertical distance $h=180$ nm. Notice that for this $h$, the emitter-metal distance is about $47$ nm, which is comparable to the optimum distances found in other PW geometries like metallic wires or wedges \cite{Martin-Cano10}. As shown in Fig. \ref{stationary-laser} (dotted blue line), when the system is prepared in the symmetric state $\ket{+}$ by a large spot laser which excites equally the two qubits ($\Omega_1=\Omega_2$), $C_\infty$ gets its maximum value for $d$ close to an odd multiple of  $\lambda_{pl}/2$, a situation in which $\gamma_{12}$ is closer to $-\gamma$ and the singlet state $\ket{+}$ is almost decoupled from the cascade decay involving the other three states (see Fig. \ref{scheme}(b)). One can also prepare the state $\ket{-}$ by pumping the two qubits with the same frequency and same intensity but phase-shifted by $\pi$, which is equivalent to using $\Omega_2=-\Omega_1$. As expected, the result is just shifted with respect to the previous case: $C_\infty$ gets its maximum value for  $d$ being an even multiple of  $\lambda_{pl}/2$ (dashed red curve in Fig.4). The $\lambda_{pl}$ periodicity of these two previous cases changes to $\lambda_{pl}/2$ when one of the two  qubits is privileged with respect to the other by focusing the laser beam only on it ($\Omega_1 \gg \Omega_2$). Here we present the case $\Omega_1=0.15 \gamma$  while $\Omega_2=0$ (see solid black curve in Fig. 4). Notice that, in the three cases displayed in Fig. 4(a), large values for the concurrence at distances larger than the operating wavelength are attainable when using realistic values for both $\beta$ and $L$. 

The robustness of our proposal with respect to changes in the $\beta$-factor is analyzed in Fig. 4(c). As expected, the steady-state concurrence obtained at $d=\lambda_{pl}$ is reduced when $\beta$ is smaller than $1$. However, this change is not very abrupt, as the concurrence is decreased by a factor of $2$ when $\beta$ is reduced from $1$ to $0.8$. Note that the scheme presented here for PWs could be also operative in other types of waveguides that display large $\beta$-factors as, for example, photonic crystal waveguides \cite{Lalanne,Lodahl}. 

In conclusion, plasmon polaritons in realistic one-dimensional waveguides are excellent candidates to act as mediators for achieving large values 
of entanglement between two distant qubits. We have shown that the large $\beta$-factors associated with these waveguides and the fact that the coherent and incoherent parts of the two-qubit coupling driven by plasmons are phase-shifted allow to modulate populations and correlations between quantum states. We believe that our findings could also be useful for implementing the concept of dissipative engineering of states \cite{verstraete09a,krauter10a,alharbi10a}.

A.G.-T. and D.M.-C. contributed equally to this work.
Work supported by the
Spanish MICINN (MAT2008-01555, MAT2009-06609-C02, CSD2006-00019-QOIT and CSD2007-046-
NanoLight.es) and CAM
(S-2009/ESP-1503). A.G.-T. and D.M.-C acknowledge FPU grants (AP2008-00101 and AP2007-00891, respectively) from the Spanish Ministry of Education.

\bibliography{plasmon}

\end{document}